\documentclass[aip,
rsi,%
 amsmath,amssymb,
 reprint,%
]{revtex4-1}

\usepackage{graphicx}
\usepackage{dcolumn}
\usepackage{bm}
\usepackage{mhchem}
\usepackage{gensymb}
\usepackage{textcomp}
\begin{document}


\title{High critical magnetic field superconducting contacts to Ge/Si core/shell nanowires}

\author{Z. Su}
 \affiliation{Department of Physics and Astronomy, University of Pittsburgh, Pittsburgh, PA 15260, USA}
 
\author{A. Zarassi}
\affiliation{Department of Physics and Astronomy, University of Pittsburgh, Pittsburgh, PA 15260, USA}

\author{B. M. Nguyen}
\affiliation{Center for Integrated Nanotechnologies, Los Alamos National Laboratory, Los Alamos, NM 87545, USA}

\author{J. Yoo}
\affiliation{Center for Integrated Nanotechnologies, Los Alamos National Laboratory, Los Alamos, NM 87545, USA}

\author{S. A. Dayeh}
\affiliation{Department of Electrical and Computer Engineering, University of California, San Diego, La Jolla,
CA 92037, USA}
\affiliation{Graduate Program of Materials Science and Engineering, University of California, San Diego, La Jolla,
CA 92037, USA}
\affiliation{Department of NanoEngineering, University of California, San Diego, La Jolla, CA 92037, USA}

\author{S. M. Frolov}
 \email{frolovsm@pitt.edu}
\affiliation{Department of Physics and Astronomy, University of Pittsburgh, Pittsburgh, PA 15260, USA}

\date{\today}

\begin{abstract}
Contacts between high critical field superconductors and semiconductor nanowires are important in the context of topological quantum circuits in which superconductivity must be sustained to high magnetic fields. Here we demonstrate gate-tunable supercurrent in NbTiN-Ge/Si core/shell nanowire-NbTiN junctions. We observe supercurrents up to magnetic fields of 800 mT. The induced soft superconducting gap measured by co-tunneling through a quantum dot is $220$ $\mu$eV. To improve contact transparency,
we deposit an aluminum interlayer prior to NbTiN and observe a systematic change in the device pinch-off voltages with aluminum thickness. We inform future multi-step device fabrication by observing that aluminum anneals
with the Ge/Si nanowire at 180\,\celsius, the baking temperature of common electron beam lithography resists.
\end{abstract}

\maketitle

Josephson junctions based on semiconductor nanowires (NWs) have become a fertile research platform in recent years. Charge tunanbility of semiconductors has been used to implement supercurrent transistors~\cite{supercurrent_transistors_Delft}, Josephson $\pi$-junctions~\cite{Jordan_von_Dam_2006} and Cooper-pair beam splitters~\cite{Schonenberger_Nature}. Theoretical proposals for realizing Majorana zero modes in semiconductor-superconductor hybrid structures~\cite{Lutchyn PRL 2012, Oppen PRL 2012} have led to a series of experiments aiming at the elucidation of Majorana fermions~\cite{delft majorana, marcus majorana}. These  works focused on nanowires with strong spin-orbit interaction, large g-factors, ballistic transport along the NW, and induced superconductivity in the NW.

Ge/Si core/shell NWs were proposed as candidates for Majorana studies, which requires high mobility, strong spin-orbit interaction, large g-factors and induced superconductivity~\cite{D. Loss Majorana PRB}. Observation of one-dimensional hole gas~\cite{1D gas PANS} suggests (quasi-)ballistic transport in junctions made in these NWs. Strong spin-orbit interaction has been predicted~\cite{Loss ge/si SOI} and pointed at by several experiments~\cite{hole spin relaxation,antilocalization:prl2014}; enhanced $g$-factors were reported~\cite{hole spin relaxation, g-factor 2016}. Supercurrents were observed in Al-Ge/Si-Al junctions~\cite{Lieber supercurrent}. However, pure aluminum is not suitable for future development of advanced topological quantum circuits based on Majorana fermions due to the low bulk critical magnetic field of aluminum. 

In this paper, we report superconductivity in Ge/Si NWs induced by niobium titanium nitride alloy contacts. NbTiN has a high critical magnetic field of more than 10 Tesla. We demonstrate the Josephson supercurrent through the NbTiN-Ge/Si-NbTiN junctions and investigate its magnetic field behavior. We evaluate the induced gap in the nanowire through tunneling in the pinched-off regime of the junction. We find that a Ti or Al interlayer between the Ge/Si nanowire and NbTiN improves the contact transparency. We investigate the effect of the Al interlayer on the nanowire device pinch-off voltage, and report the effect of aluminum alloying with the nanowire. 

\begin{figure}[t]
  \includegraphics[width=8.5cm]{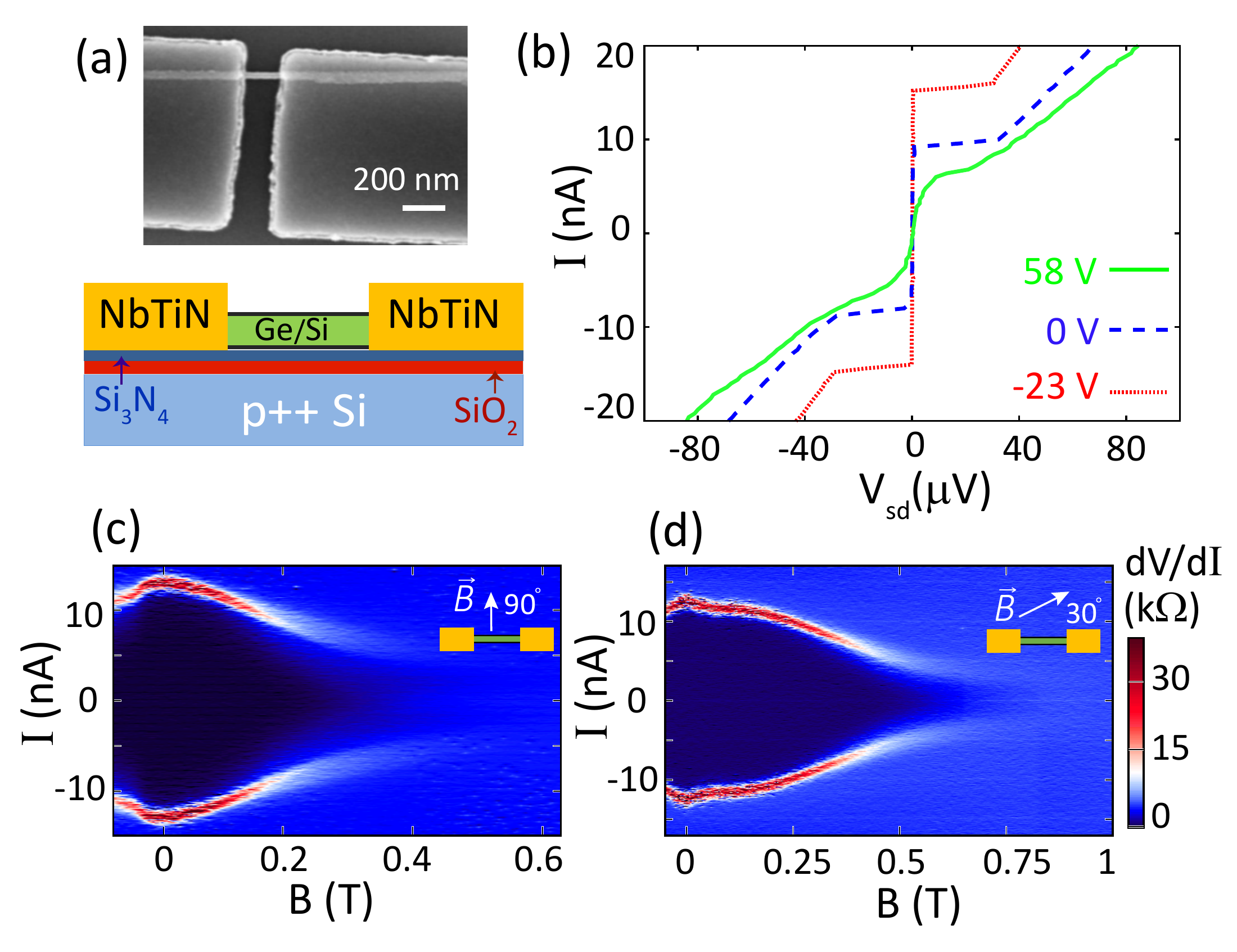}
  \caption{(a) (Top) Scanning electron micrograph of a Ti/NbTiN-Ge/Si-Ti/NbTiN device. (Bottom) Side view schematic of the device. (b) IV characteristics of device A at $V_{bg} = -23$ V (red), $0$ V (blue) and $58$ V (green). (c-d) The differential resistance $dV/dI$ as a function of current bias and magnetic field for device A at $V_{bg} = -10$ V (panel (c)) and device B at $V_{bg} = 0$ V (panel (d)).  In-plane magnetic field orientations are indicated by cartoons.}
\label{fig1}
\end{figure}

The devices are fabricated on doped Si substrates which serve as back gates, and are covered by thermal \ce{SiO2} ($285$ nm) and chemical vapor deposited \ce{Si3N4} (50 nm) (Fig.\ref{fig1}(a)). Ge/Si nanowires
~\cite{shadi1,shadi2,shadi3} with core diameters of 20-40 nm and shell thickness of 2 nm~\cite{minh nw} are positioned on the substrates using a micromanipulator~\cite{manipulation2011}. The superconducting leads are patterned by electron beam lithography with nominal spacing of 200 nm followed by magnetron sputtering of NbTiN. A 3 nm interlayer of Ti (devices A and B) or Al (device C) is deposited prior to NbTiN. Before metals deposition, a two-second buffered hydrofluoric (BHF) acid etch is applied to remove the NW surface native oxide. \textit{In-situ} argon plasma cleaning yielded higher contact resistances than BHF cleaning. We perform transport measurements in a dilution refrigerator with the base temperature of 40 mK. 

A typical current-voltage (IV) characteristic is shown in Fig.\ref{fig1}(b). Because the charge carriers in Ge/Si NWs are holes, a more negative back gate voltage $V_{bg}$ leads to a lower normal state resistance ($R_n$) and a higher switching supercurrent ($I_c$). The excess current ($I_{exc}$) due to Andreev reflection extracted from the IV traces is in the range of $1-4$ nA. Contact transparency is estimated using the Blonder-Tinkham-Klapwijk theory from the ratio $eI_{exc}R_n/\Delta_{in}\sim0.05$, where $e$ is the elementary charge. This corresponds to a transparency of $\sim15\%$ in the short junction limit~\cite{BTK prb}. Note that this is a lower bound estimate on the contact transparency since the ratio above is reduced for finite length junctions~\cite{InAs H.Q. Xu}. On the other hand, the $eI_{c}R_{n}/\Delta_{in}$ ratio is 0.2 (the ratio to bulk gap is 0.03, given a NbTiN critical temperature $T_c = 11$ K) which is lower than the theoretical limit of $\pi$. Limited contact transparency and premature switching out of the supercurrent state caused by thermal activation can contribute to the reduction of the $I_cR_n$ product\cite{tinkham}. The fact that supecurrent in Ge/Si nanowires is carried by holes which have a different momentum $J$ than electrons in the lead superconductor may also contribute to this reduction, though this aspect is so far unexplored.

Magnetic field evolution of supercurrents in two devices is presented in Figs.\ref{fig1}(c,d). The switch out of the supercurrent state and into the finite voltage state corresponds to a peak in $dV/dI$. It is clearly visible up to $400$ mT but can also be traced to higher fields. Both devices demonstrate monotonous decrease in $I_c$ as magnetic field is increased. The fact that supercurrent survives to higher fields in device B is consistent with the field aligned closer to the nanowire axis in this device, which reduces the magnetic flux threading the junction area for a given field. No nodes or oscillations in the switching current are observed up to fields of 1 Tesla, throughout the range of resolved supercurrent features. Assuming purely Fraunhofer-type interference within the junction only (zero field in the leads), the first node is expected at 300 mT for device A and 600 mT for device B, though other geometrical factors such as the circular nanowire cross-section and Meissner effect may contribute to the absence of the nodes. The semiconductor nanowires are tuned to the multi-mode regime with several transverse one-dimensional subbands occupied, a factor which favors critical current oscillations due to inter-mode interference influenced by magnetic field.

\begin{figure}[t]
\includegraphics[width=8.5cm]{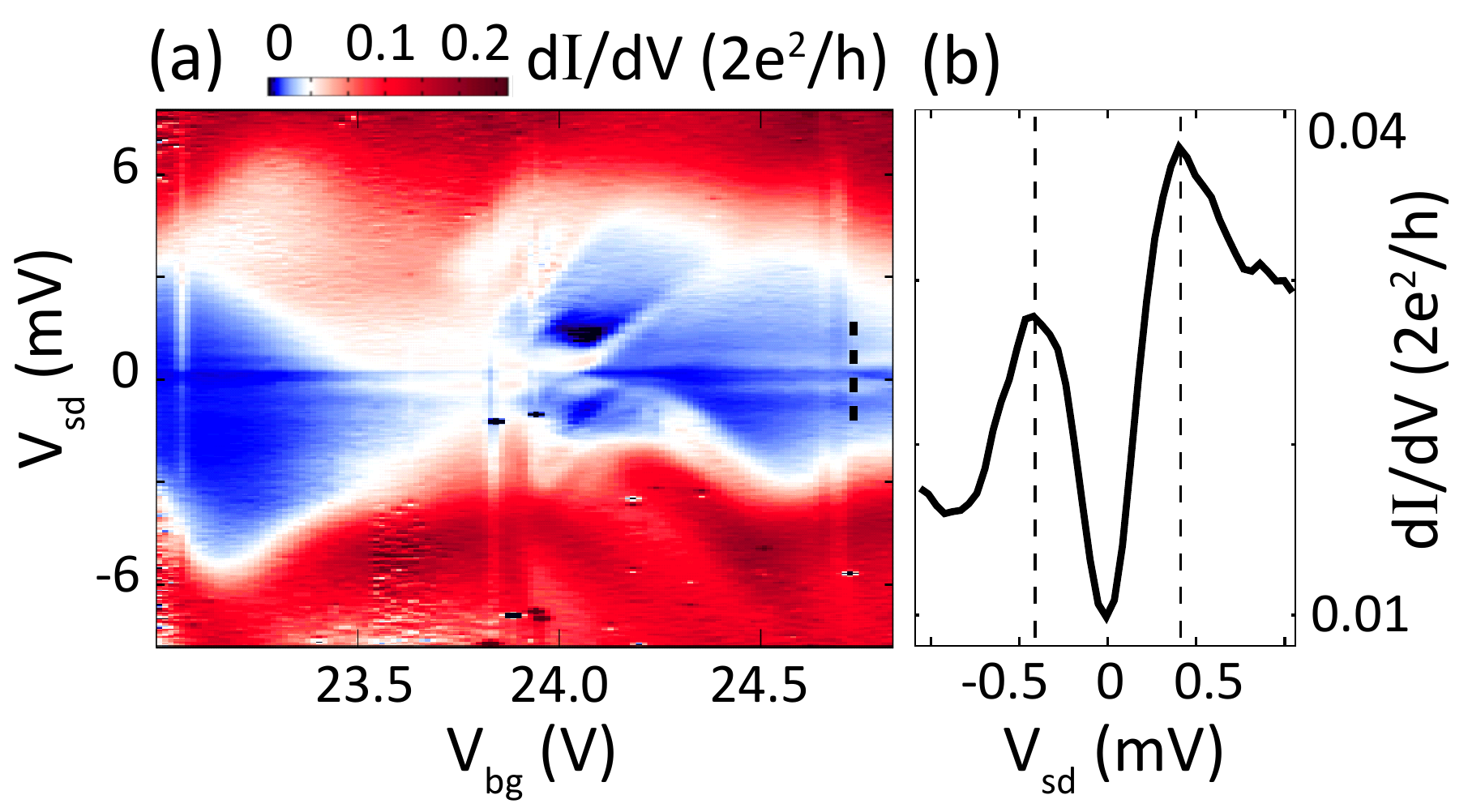}
\caption{ (a) Coulomb diamond measurements on an accidental quantum dot in device C in the near pinch-off regime, a dashed line marks a line cut shown in panel (b). (b) A line cut showing $dI/dV$ as a function of the bias voltage. Peaks separated by $4\Delta_{in}$ are marked by dashed lines.}
 \label{fig2}
\end{figure}

As part of evaluating the prospects of hybrid junctions studied here for the realization of Majorana zero modes, we measure the induced superconducting gap $\Delta_{in}$. The gap is studied in the co-tunneling regime through unintentional quantum dots formed in the junctions near pinch-off~\cite{co-tunneling}. Fig.\ref{fig2}(a) shows differential conductance in device C as a function of back gate voltage. The boundaries of Coulomb diamonds are broadened due to co-tunneling, as the quantum dot barriers remain low. We observe a pair of sharp horizontal resonances of high differential conductance located symmetrically around zero bias at $V_{sd}=2\Delta_{in}=\pm ~440$ $\mu V$. 
At biases in between the peaks conductance is suppressed but remains non-zero. Thus the induced superconducting gap here is of the so-called ``soft gap'' type (see Fig.\ref{fig2}(b)). This is not ideal for topological quantum circuit applications since any subgap conductance obfuscates signatures of Majorana fermions and leads to topological qubit decoherence. Possible causes of the soft gap are disorder and quasiparticle poisoning~\cite{stanescu_softgap,takei_softgap}. The soft gap observed here is of the same magnitude as in an early Majorana study based on InSb nanowires~\cite{delft majorana}. Since the same NbTiN alloy was used in both cases, the soft gap may be related to the properties of the NbTiN film, such as its granularity and alloy disorder. Relatively small interface transparency of devices studied here can also be a contributing factor.

\begin{figure}[t]

\includegraphics[width=8.5cm]{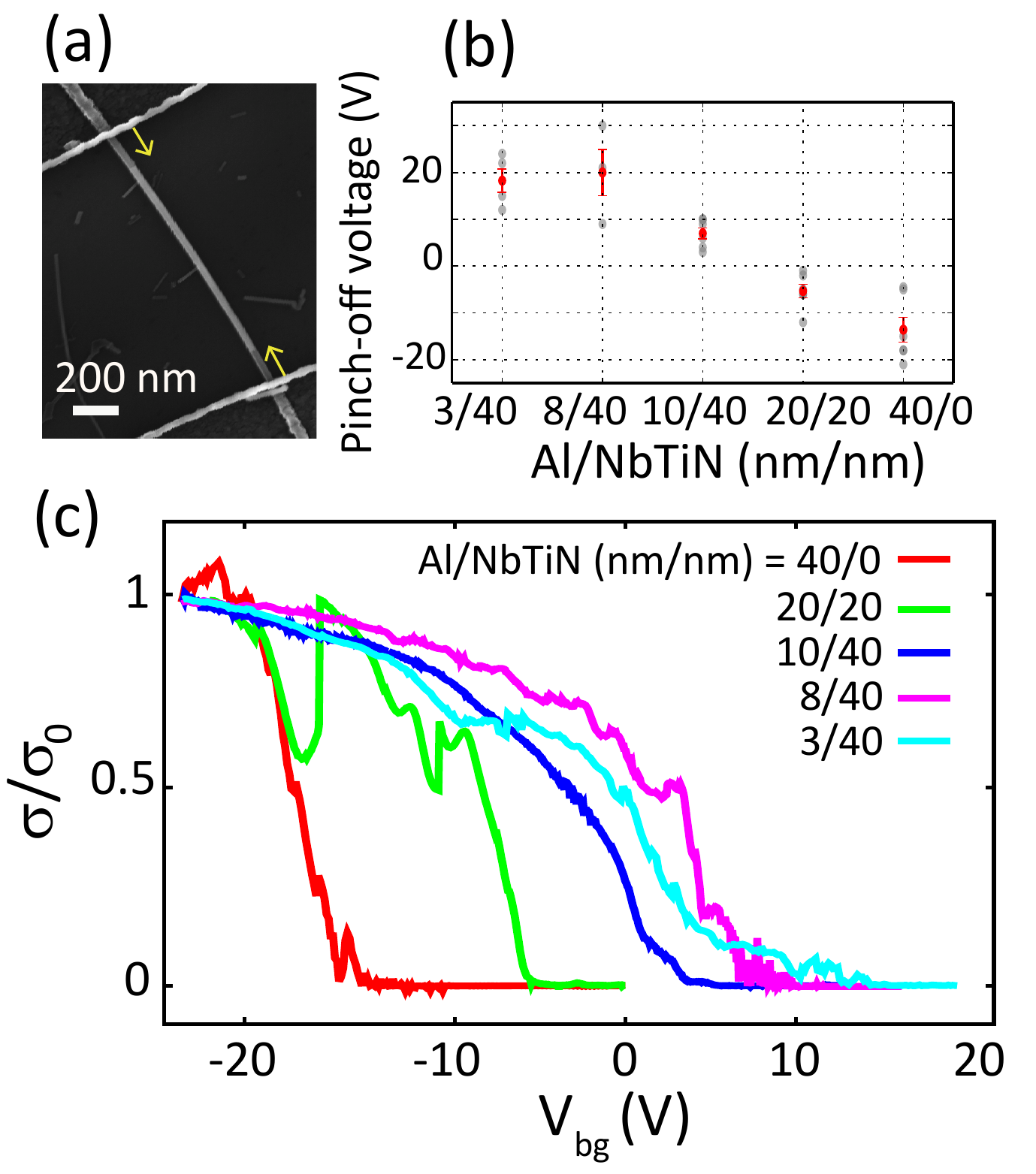}
\caption{(a) Al-Ge/Si-Al device after a rapid annealing at 180\,\celsius. The arrows indicate alloyed regions of the nanowire.
(b) Pinch-off voltages of 21 devices with various Al/NbTiN thicknesses. The average value for each Al/NbTiN thickness combination is indicated by the red dot. (c) Typical pinch-off traces of devices for various Al/NbTiN thickness combinations. The conductance is normalized by conductance $\sigma_0$ measured for each device at $V_{bg}=-24$ V.}
 \label{fig3}
\end{figure}

To improve the yield of transparent contacts we explore contact annealing in the presence of forming gas ($5\%$~\ce{H2} and $95\%$~\ce{N2} ) at 1 bar. For NbTiN contacts with thin Al or Ti interlayers (3 nm), no measurable change in the saturation resistance is observed after annealing at temperatures up to 400\,\celsius. However, we observe that pure Al contacts alloy rapidly, within seconds, at temperatures as low as 180\,\celsius. Fig.\ref{fig3}(a) demonstrates the alloying of 200 nm regions next to each contact. Saturation junction resistances are reduced by more than one order of magnitude after the annealing and low resistance ($\leq 10~k \Omega$) devices are obtained with a relatively high yield of $\sim1/4$ for pure Al contacts. It is important to be aware of the Al-alloying effect because the annealing temperature is close to the temperature used during standard nanofabrication. For example, electron beam resist PMMA is baked at 180\,\celsius, so if top gates are fabricated after pure aluminum contacts, those contacts are going to be annealed with the nanowires.

We find that the pinch-off voltages of the junctions can be modified by changing the thickness of the Al interlayer. Figs.\ref{fig3}(b),(c) show the low temperature pinch-off voltage data from many unannealed devices with different Al and NbTiN layer thicknesses. Devices based on NbTiN with a thin Al interlayer  have large positive pinch-off voltages. The increase of Al thickness reduces the pinch-off voltages until ultimately the pinch-off voltage becomes negative. The pure Al contact results on average in negative pinch-off voltages in these devices. 

The pinch-off voltages can also be affected by the interface between the NW and the gate dielectric. We observe that the pinch-off voltages tend to shift to more positive values when \ce{HfO2} is used instead of \ce{Si3N4}. Combined effects of varying dielectric and the contact stack offer a pathway to adjusting the working point of a device to be close to zero gate voltage, which is expected to minimize charge noise.

In summary, we developed superconducting contact recipes for Ge/Si NWs based on NbTiN, a superconductor with a high critical magnetic field. Josephson junction devices based on these recipes exhibit gate-tunable Josephson current which can survive under magnetic fields up to $800$ mT. An induced superconducting gap of $220$ $\mu eV$ is measured, although the sub-gap conductance remains significant. Our systematic fabrication and transport measurement studies show that increasing the Al interlayer thickness modifies the pinch-off voltages of the devices dramatically, from positive to negative values, and that contacts with thicker Al layers can be annealed in order to obtain transparent contacts with higher yields. The present results indicate promise for a future realization of Majorana zero modes in this materials system, although further studies are required to achieve ``hard'' induced gaps~\cite{superhard gap}. 

The Ge/Si nanowire growth was performed at the Center for Integrated Nanotechnologies (CINT), U.S. Department of Energy, Office of Basic Energy Sciences User Facility at Los Alamos National Laboratory (Contract DE-AC52-06NA25396) and Sandia National Laboratories (Contract DE-AC04-94AL85000). We thank E. Lee, S. De Franceschi, V. Bouchiat and M. Hocevar for useful discussions. S.A.D. acknowledges NSF support under DMR-1503595 and ECCS-1351980. S.M.F. acknowledges NSF DMR-125296, ONR N00014-16-1-2270 and Nanoscience Foundation, Grenoble.


\end{document}